\documentstyle[12pt]{article}
\input epsf
\begin{document}
\pagestyle{plain}
\pagenumbering{arabic}
\date{January 20, 1997}
\baselineskip=12pt
\begin{center}
\begin{large}
{\bf Diffusive Evolution of Stable and Metastable Phases II:
     Theory of Non-Equilibrium Behaviour in Colloid-Polymer Mixtures} \\
\end{large}  
\bigskip
{\em R. M. L. Evans, \\ \smallskip W. C. K. Poon} \\
\medskip
Department of Physics and Astronomy, \\
The University of Edinburgh, \\
JCMB King's Buildings, \\
Mayfield Road, Edinburgh EH9 3JZ, UK \\
e-mail: r.m.l.evans@ed.ac.uk \\
\hspace{0.7cm} w.poon@ed.ac.uk \\
\bigskip
\medskip
{\bf Abstract }
\end{center}

\begin{quote} By analytically solving some simple models of
phase-ordering kinetics, we suggest  a mechanism for the onset of
non-equilibrium behaviour in colloid-polymer mixtures. These mixtures can
function as models of atomic systems; their physics therefore
impinges on many areas of thermodynamics and phase-ordering. An exact
solution is found for the motion of a single, planar interface
separating a growing phase of uniform high density from a
supersaturated low density phase, whose diffusive depletion drives the
interfacial motion. In addition, an approximate solution is found for
the one-dimensional evolution of two interfaces, separated by a slab of a
metastable phase at intermediate density. The theory predicts a critical
supersaturation of the low-density phase, above which the two interfaces become
unbound and the metastable phase grows {\em ad infinitum}. The growth of the
stable phase is suppressed in this regime. \end{quote}



\baselineskip=16pt

\section{Introduction}

Metastability and phase transition kinetics in condensed matter are
rapidly expanding fields of research \cite{Gunton83,Bray94}.
Experimentally, it is
increasingly realised that {\it colloidal systems} exhibit unique
properties which make them attractive candidates for studying these
topics. Advances in synthetic chemistry mean that suspensions of
particles of well-characterized shapes and sizes, and with precisely
tunable interparticle interactions, can be routinely produced. The {\it
equilibrium} phase behaviour of such model colloids can be studied
experimentally and understood in some detail using the standard tools
of statistical mechanics \cite{Poon95a}. Moreover, colloids are much
larger than atoms, the typical dimension of a colloid being $L \approx
0.5 \mu m$, and the solids they form (crystal, glass, gel) are
mechanically weak, a typical modulus being $G \sim k_{B}T/L^{3} \approx
10^{-2}Nm^{-2}$. Thus, a colloidal crystal can be `shear melted' to the
metastable colloidal fluid state simply by shaking. Structural dynamics
of colloidal systems are also slow, the scale being set by the time a
free particle takes to diffuse its own dimension, $\tau_{R} \sim
L^{2}/D$. Estimating the diffusion coefficient $D$ by the
Stokes-Einstein relation for a spherical particle of radius $L \sim 0.5
\mu m$ suspended in a liquid of viscosity $\eta \sim 10^{-3} Nm^{-2}s$,
we get $D = k_{B}T/6\pi\eta L \sim 4 \times 10^{-13} m^{2}s^{-1}$, so
that the characteristic relaxation time in colloidal systems is
$\tau_{R} \stackrel{>}{_{\sim}} 1\,s$. Thus, the crystallization of a
metastable colloidal fluid can take minutes, hours or even days. These two
features, the ease with which metastable states can be prepared and the
long relaxation times involved, mean that the study of phase transition
kinetics and metastability in model colloids is gaining increasing
attention. For example, there is a growing literature on
crystallization kinetics in nearly-hard-sphere PMMA colloids \cite{He96}.
The glass transition in the same system is now recognized as being one
of the simplest examples of that phenomenon, and has become a testing
ground for sophisticated theories \cite{vanMegen94}.

In recent work aimed at elucidating the equilibrium phase behaviour of
mixtures of nearly-hard-sphere colloids and non-adsorbing polymers, we
have also observed a rich variety of non-equilibrium behaviour
\cite{Pusey93,Poon95b}. The well-characterized nature of our experimental
system, coupled with detailed data from small-angle light scattering
\cite{Poon95b}, have allowed us to suggest a possible link between the
onset of formation of non-equilibrium phases and the presence of a `hidden',
metastable minimum in the free energy landscape. In this paper, we
review the experimental evidence leading to this suggestion, and model
the kinetic consequences of such a three-minima free energy landscape
in one dimension using a continuum approach. We conclude by pointing to
the possible applications of this model in other `soft' systems. Some of our
work relies on results from a companion paper \cite{Evans97a}; the main ideas
are summarized in \cite{Evanslet}.

\section{Non-Equilibrium Behaviour in Colloid-Polymer Mixtures}
\label{behaviour}

Free polymer alters the phase behaviour of colloids via the depletion mechanism
\cite{Lekkerkerker92}. Exclusion of polymer from the region between the surfaces
of two nearby colloidal particles gives rise to an unbalanced osmotic
pressure pushing the particles together, resulting in an attractive
depletion potential, $U_{\rm dep}$. In the case of hard-sphere-like colloids,
the total interparticle interaction \cite{Ilett95} in the presence of polymers
is given approximately by
\begin{equation}
  U(r)  = \left\{ \begin{array}{ll}
        + \infty & \mbox{for $r \leq \sigma$}  \\
        - \Pi_{p}(\mu_{p})V_{\rm overlap} = U_{\rm dep}
                                &\mbox{for $\sigma < r \leq
\sigma+2r_{g}$} \label{eq:udep}\\
        0 & \mbox{for $r > \sigma+2r_{g}$} 
        \end{array}
        \right. 
\end{equation}
where $\sigma = 2a$ is the particle diameter and $\Pi_{p}(\mu_{p})$ is the
osmotic pressure of the polymer as a function of the polymer chemical
potential, $\mu_{p}$. $V_{\rm overlap}$ is the volume of the
overlapping depletion zones between two particles at an inter-centre
separation of $r$. Explicitly
\begin{equation}
  V_{\rm overlap} = \left\{1 - \frac{3r}{2\sigma (1+\xi)} +
   \frac{1}{2}\left[\frac{r}{\sigma(1+\xi)}\right]^{3}\right\} 
   \times \frac{\pi}{6} \sigma^{3} (1 + \xi)^{3},
\label{eq:overlap} 
\end{equation} 
where $\xi = r_{g}/a$ denotes the ratio of the radius of
gyration of the polymer, $r_{g}$, to the radius of the colloidal
particle, $a$.

\begin{figure}
  \epsfxsize=14cm
  \begin{center}
  \leavevmode\epsffile{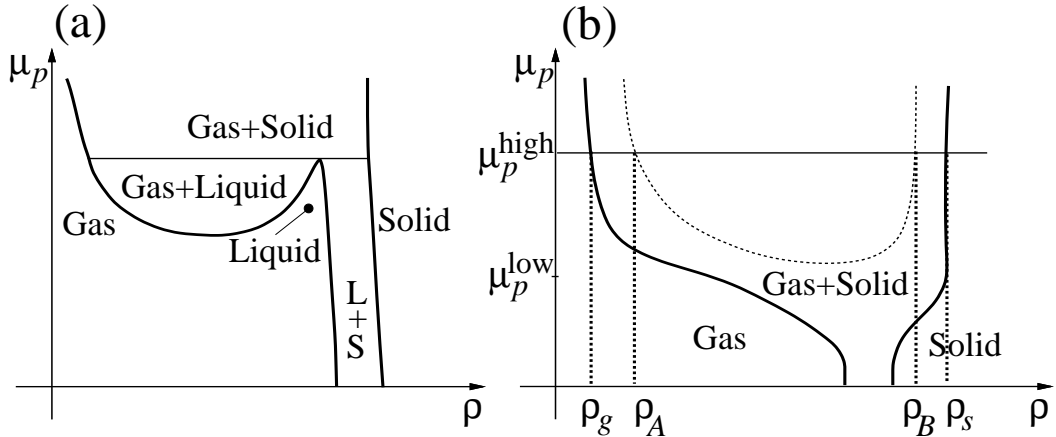}
  \caption{Phase diagrams of colloid-polymer mixtures in the colloid
   concentration ($\rho$)--polymer
   chemical potential ($\mu_{p}$) plane, for size ratio $\xi$ (a) above
   and (b) below $\xi_{\rm crit}$. The dotted curve is a ``hidden" gas-liquid
   binodal. For (b), the free energy plots for two
   polymer chemical potentials, $\mu_{p}^{\rm low}$ and $\mu_{p}^{\rm high}$,
   are shown in figure \protect\ref{fe}a and b respectively. The values of
   $\rho$ for the various boundaries at $\mu=\mu_{p}^{\rm high}$ are given by
   the common tangent construction in figure \protect\ref{fe}b.}
  \label{pd}
  \end{center}
\end{figure}

Theory \cite{Lekkerkerker92} predicts and experiments \cite{Ilett95} show that
the phase diagram of a mixture of hard-sphere colloids and
non-adsorbing polymer is a function of the size ratio, $\xi$. The phase
diagram at large enough $\xi$ is reminiscent of that of a simple atomic
substance, figure \ref{pd}a, with the polymer chemical potential
($\mu_{p}$) playing the role of an inverse temperature. (Temperature itself is
not an axis of the phase diagram because, in systems with only excluded-volume
interactions, no energy scale exists \cite{Lowen94}.) Colloidal
gas, liquid and crystal phases are possible, with regions of binary
coexistence between pairs of these phases and triple coexistence at a
particular value of $\mu_{p}$. As the size ratio $\xi$ is decreased,
the region of colloidal gas-liquid coexistence shrinks until, below a
critical value $\xi_{\rm crit}$, the gas-liquid coexistence region
disappears altogether from the equilibrium phase diagram, figure \ref{pd}b;
now, only colloidal fluid or crystal phases exist in equilibrium.
Experimentally, $\xi_{\rm crit} \approx 0.25$.

Focus now on the phase diagram for the case of $\xi < \xi_{\rm crit}$,
figure \ref{pd}b. Experiments confirm that moderate concentrations of a
small non-adsorbing polymer cause a suspension of hard spheres to phase
separate into coexisting colloidal fluid and crystal phases. At higher
colloidal concentrations, or higher $\mu_{p}$, across a {\it
non-equilibrium boundary} (NEB), crystallization is suppressed
\cite{Pusey93}, \cite{Poon95b}. Instead, a variety of non-equilibrium
aggregation behaviour is observed. Detailed small-angle light scattering
studies \cite{Poon95b} have allowed the classification of the kinetic
behaviour above the NEB into three types. Just across the NEB, the
behaviour is `nucleation like'. This is characterized by an initial
latency period consistent with nucleation, after which dense, amorphous
droplets separate out, forming an amorphous sediment which begins to
crystallize only on a much longer time-scale. The
`nucleation-like' regime is of primary interest to us in this article.
At higher $\mu_{p}$, or higher colloid concentration, the behaviour
becomes `spinodal-like', switching finally to a `transient gelation' regime
for the densest systems.

\begin{figure}
  \epsfxsize=14cm
  \begin{center}
  \leavevmode\epsffile{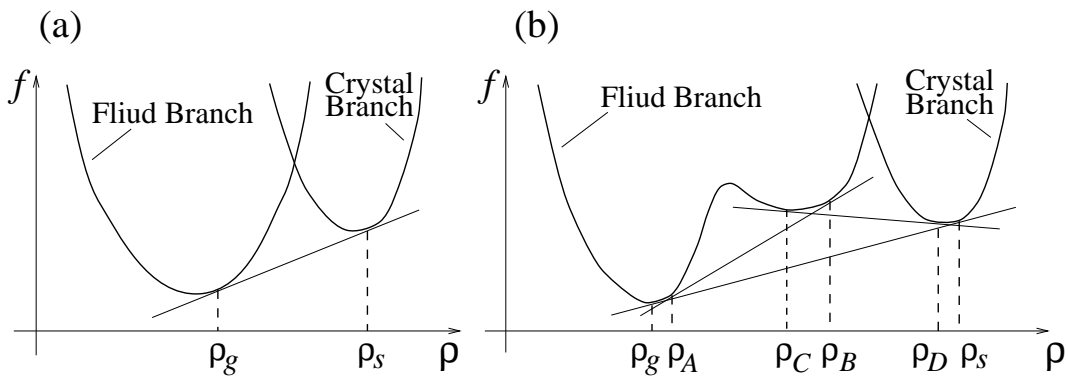}
  \caption{Fluid and crystal branches of the free energy density versus
   colloid volume fraction (a) at low polymer chemical potential (e.g.\
   $\mu_{p}^{\rm low}$ in figure \protect\ref{pd}b), when the fluid branch has
   a single minimum, and (b) at higher polymer chemical potential (e.g.\
   $\mu_{p}^{\rm high}$ in figure \protect\ref{pd}b), when two minima exist in
   the fluid branch, corresponding to gas and metastable liquid phases. The
   common tangent construction is used to find the densities of the coexisting
   equilibrium gas and solid phases $(\rho_{g},\rho_{s})$, and the metastable
   gas-liquid and liquid-solid binodal points $(\rho_{A},\rho_{B})$,
   $(\rho_{C},\rho_{D})$; see figure \protect \ref{pd}b.}
  \label{fe}
  \end{center}
\end{figure}

It should be stressed that the experimentally observed NEB is sharp and
highly reproducible. In fact, it has been suggested by one of us
\cite{Poon95b} that this boundary is given by a `hidden' gas-liquid
binodal. Following the theoretical approach in \cite{Lekkerkerker92},  we
write the free energy, $F(\rho,\mu_{p})$, of a colloid-polymer mixture
as a function of the colloid volume fraction (or `density'), $\rho$, and the
polymer chemical potential, $\mu_{p}$. $F$ can be calculated within a
mean-field framework for a disordered arrangement of colloids and
polymers, the `fluid branch', and an ordered arrangement of colloids
with polymers randomly dispersed, the `crystal branch'. At low polymer
chemical potentials, the fluid and crystal branches each show a single
minimum (figure \ref{fe}a). This gives rise to single-phase fluid
($\rho<\rho_{g}$), fluid-crystal coexistence ($\rho_{g}<\rho<\rho_{s}$), or
single-phase crystal ($\rho>\rho_{s}$). The colloid
concentrations in coexisting fluid and crystal phases are obtained by
the common tangent construction (see, e.g., \cite{DeHoff93}).

At higher polymer chemical potentials, however, the fluid branch of the
free energy shows a `double minimum' structure (figure \ref{fe}b). At
larger polymer to colloid size ratios, this double minimum can give
rise to a region of colloidal gas-liquid coexistence in the equilibrium
phase diagram, figure \ref{pd}a \cite{Lekkerkerker92,Ilett95}. For
small polymers, however, the theory predicts only separation into fluid
and crystal phases. Nevertheless, the `metastable gas-liquid binodal',
which is `buried' within the fluid-crystal  coexistence region
predicted by equilibrium thermodynamics, can still be traced out (figure
\ref{pd}b). The
equilibrium phase boundary and the metastable gas-liquid binodal,
calculated using the theory in \cite{Lekkerkerker92}, compare well with the
experimental equilibrium fluid-crystal coexistence boundary and the 
non-equilibrium boundary respectively \cite{Poon95b}, giving rise to
the speculation that the suppression of crystallization, which is the
predicted, equilibrium phase behaviour, is linked to the presence of
the hidden gas-liquid binodal.

  In this and a preceding, companion paper \cite{Evans97a}, we provide a
theoretical basis for this speculation. We model the effects of a metastable 
third minimum in the free energy curve on the kinetics of phase ordering,
focusing on post-nucleation kinetics. A system which has been homogenized
(e.g.\ by shear-melting) to some uniform density between $\rho_{g}$ and
$\rho_{s}$ (see figure \ref{fe}) must evolve towards the equilibrium state, by
separating into regions of density $\rho_{g}$ and $\rho_{s}$. If $f(\rho)$ has
positive curvature at the initial density, then the evolution begins by
nucleation, whereby a thermal fluctuation causes part of the system to cross
the barrier between the two stable wells in $f$. Once such a nucleus has formed
(in the {\em early} stage of phase ordering), it grows by diffusively depleting
the surrounding, supersaturated medium (the {\em growth} stage
\cite{Tokuyama93}).

  In the previous
paper \cite{Evans97a}, we consider the implications for short range (on the
length scale of the transition zone between neighbouring phases) density
variations and stability, in the context of quasi-steady-state motion. There,
and in Ref.\ \cite{Evanslet},
it is shown that an interface between regions of different densities may split,
in the presence of a metastable minimum, into two parts. The purpose of the
present article is to model the evolution of a split interface during the
growth stage, considering large length scales (so interfaces appear as sharp
steps, and relatively gentle concentration gradients control the diffusive
dynamics) and time-scales (so motion is time-dependent, not steady-state). We
will find that growth dynamics involving split interfaces contrasts greatly
with normal growth behaviour.

  The plan of this paper is as follows. The model introduced in section
\ref{exact}, involving a nucleus with an ordinary, unsplit interface, sets the
scene for the methods employed. It is a completely soluble model for the motion
of a single domain wall. In section \ref{competing}, we analyse a more elaborate
scenario, where two domain walls (the two parts of a ``split" interface) enclose
a region of metastable density, and
compete for the flux of condensing material. This model exhibits behaviour
consistent with the non-equilibrium boundary in the colloid-polymer phase
diagram. The limitations of our work are discussed in section \ref{discussion},
while applications to related systems are given in section \ref{related}.

  In the models presented, growth is limited by diffusion of matter only. This
simplification holds in many soft condensed systems, for which diffusion of
latent heat is irrelevant since the entropy is dominated by the degrees of
freedom of the solvent \cite{Lowen94}. Moreover, we assume that the system may
be described by a {\em single} conserved order parameter (such as the density
of colloidal particles). Formally this does not apply in colloid-polymer
mixtures, where the polymer concentration is a second conserved quantity.
However, it should be a reasonable model for small $\xi$, such that the
polymers diffuse much more rapidly than the colloids. Non-conserved order
parameters describing crystallinity are also omitted, under the assumption that
such variables relax quickly compared with the diffusive order parameter, and
are therefore not rate-limiting.

\section{Exact Solution of Diffusion Equation for Two-Phase Separation}
\label{exact}

  In this section, we find an exact solution of a problem pertinent to
the intermediate-to-late-stage evolution of a two-phase system. It is
well known \cite{Tokuyama93} that, when a dense droplet grows into a
surrounding, supersaturated medium, the speed of motion of the interface
is proportional to $t^{-1/2}$. We solve an idealized one-dimensional model
for the motion of a single interface in such a situation, and find the
constant of proportionality as a function of the supersaturation of the
ambient medium. The results obtained in this section will be used in
section \ref{competing} to investigate the effect on interfacial motion
of a `metastable minimum' in the free energy landscape.

\subsection{Idealized Model}

\begin{figure}[h]
  \epsfxsize=10cm
  \begin{center}
  \leavevmode\epsffile{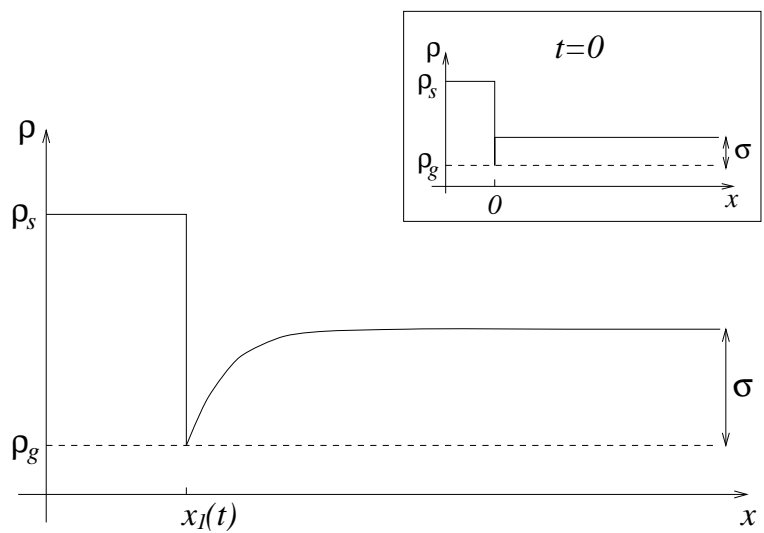}
  \caption{Graph of density
   $\rho$ against position $x$. On the left is the dense region which has
   been nucleated, and is growing at the uniform density $\rho_{s}$, which
   is its equilibrium density. The interface is located at $x=x_{1}$,
   which is a function of time, since matter is flowing down the
   concentration gradient in the sparse phase, and condensing onto the
   high-density region. At the base of the interface, the density is that
   of the coexisting gas, $\rho_{g}$. The `ambient'
   density far from the interface is $\rho_{g}+\sigma$. Inset: The initial
   ($t=0$) configuration of the system, with a delta-function singularity
   at the base of the domain wall.}
  \label{diff}
  \end{center}
\end{figure}

  Consider an infinite, one-dimensional system whose equilibrium state
is two-phase coexistence, which has been homogenized to a uniform
density ({\em e.g.\ }by shear melting). In the system, a region of the
high-density phase has nucleated at its equilibrium density, and the
interface between the high- and low-density phases has locally
equilibrated. The domain wall proceeds to move by depleting the supersaturated
low-density phase. Let the interface be located at $x=x_{1}(t)$ and let
the density of the high-density phase be uniform at $\rho=\rho_{s}$. So
the evolution of the system is solely due to the dynamics of the
low-density region. In this phase, the density is $\rho_{g}$ at the
interface ($x=x_{1}$), and tends asymptotically to the supersaturated
value $\rho_{g}+\sigma$ as $x\rightarrow\infty$. The assumption that
the density (and hence chemical potential) is fixed at the equilibrium
value (given by the double tangent construction in figure \ref{fe}a) at
the interface is physically valid if the interface moves sufficiently
slowly \cite{Evans97a}. The problem of finding $x_{1}(t)$ is that of
solving the diffusion equation
\begin{displaymath}
  \frac{\partial\rho}{\partial t} = D \frac{\partial^2\rho}{\partial x^2}
\end{displaymath}
with boundary conditions $\rho(\infty,t)=\rho_{g}+\sigma$ and
\begin{equation}
\label{BC1}
  \rho(x_{1},t)=\rho_{g}
\end{equation}
with a moving boundary $x_{1}(t)$, whose velocity depends on the flux
at that point since, by conservation of matter at the interface,
\begin{equation}
\label{BC2}
  (\rho_{s}-\rho_{g})\,\dot{x}_{1} = D \rho'(x_{1}).
\end{equation}
The initial conditions are $x_{1}=0$ and
$\rho(x\!\!>\!\!x_{1})=\rho_{g}+\sigma$, with a negative delta-function
singularity at $x=x_{1}$, since $\rho(x_{1})=\rho_{g}\;\; \forall\;\;
t$. The system is depicted, at some later time $t$, (and the initial condition
is inset) in figure \ref{diff}.

\subsection{Solution}
\label{solution}

Solving the diffusion equation with a moving boundary (a Stefan problem)
is difficult. The problem is overcome by replacing the semi-infinite region
in which the diffusion equation is to be applied, by an infinite region,
with a source or sink at $x_{1}$, whose strength $s(t)$ is such that the
region $x>x_{1}$ cannot distinguish it from a moving interface. The behaviour
in $x<x_{1}$, which is just a construct of the method, should
simply be ignored. The initial conditions in this region may be freely
chosen to facilitate the solution. Let
$\rho\left(x\!\neq\!x_{1}(0)\,,t\!=\!0\right)=\rho_{g}+\sigma$ and let
$y$ be defined as the deviation from this ambient density;
$y=\rho-\rho_{g}-\sigma$. So the equation to be solved is the diffusion
equation with a source at $x=x_{1}(t)$:
\begin{equation}
\label{eqnofmotion}
  \frac{\partial y}{\partial t} = D \frac{\partial^2 y}{\partial x^2}
   - \delta(x-x_{1}(t))\:s(t)
\end{equation}
with the wall position $x_{1}(t)$ and the construct $s(t)$ fixed by the
boundary conditions. Equation \ref{eqnofmotion} is solved by
\begin{displaymath}
  y(x,t) = -\int_{0}^{t}\!dt' \int_{-\infty}^{\infty}\!dx'\, G(x-x',t-t')\;
            \delta(x'-x_{1}(t'))\;s(t')
\end{displaymath}
using the Green's function
\begin{displaymath}
  G(x,t) = \frac{1}{\sqrt{4\pi Dt}} \exp\left(\frac{-x^2}{4Dt}\right).
\end{displaymath}
Hence the density field is given in terms of $s(t)$ and $x_{1}(t)$ by
\begin{displaymath}
  \rho(x,t) = \rho_{g}+\sigma - \int_{0}^{t}\!dt'\,
   \frac{s(t')}{\sqrt{4\pi D (t-t')}}
   \,\exp-\frac{(x-x_{1}(t'))^2}{4D(t-t')}.
\end{displaymath}
Applying equations \ref{BC1} and \ref{BC2} gives us two integral
equations in the two unknown functions $x_{1}(t)$ and $s(t)$, valid for
all positive $t$;
\begin{eqnarray}
  \int_{0}^{t}\!dt'\, \frac{s(t')}{\sqrt{4D(t-t')}}\,
    \exp - \frac{(x_{1}(t)-x_{1}(t'))^2}{4D(t-t')}
    &=& \pi^{1/2},
\label{integ1} \\
    \int_{0}^{t}\!dt'\, \frac{(x_{1}(t)-x_{1}(t'))\,s(t')}{(4D(t-t'))^{3/2}}\,
    \exp - \frac{(x_{1}(t)-x_{1}(t'))^2}{4D(t-t')}
    &=& \frac{\pi^{1/2}(\rho_{s}-\rho_{g})}{2D}\,\dot{x}_{1}(t).
\label{integ2}
\end{eqnarray}
The right-hand side of equation \ref{integ1} is independent of $t$, and hence
the $t$ dependence must be removed from the exponential in the integrand. This
requires
\begin{equation}
\label{x1}
  x_{1}(t)=a\sqrt{Dt},
\end{equation}
and hence $s(t)=b\surd(D/t)$, where $a$ and $b$ are constants in time.
Substituting into equations \ref{integ1} and
\ref{integ2} and evaluating the integrals (see appendix
\ref{integrals}) results in a closed-form expression relating the
coefficient $a$ to the relative supersaturation;
\begin{equation}
\label{answer}
  \frac{\sigma}{\rho_{s}-\rho_{g}}
   = \mbox{$\frac{\sqrt{\pi}}{2}$}\,a\,
   e^{\frac{a^2}{4}}\,\mbox{erfc}\,\mbox{$\frac{a}{2}$}
\end{equation}
where $\,\mbox{erfc}\,$ is the complementary error function, $\,\mbox{erfc}\,
x\equiv 1-\,\mbox{erf}\, x$. The coefficient $b$ is not of particular interest,
but we note that it is of the form $(\rho_{s}-\rho_{g})\times
(\mbox{function of $\frac{\sigma}{\rho_{s}-\rho_{g}}$})$.

  We have chosen to use the initial condition $x_{1}(0)=0$ in deriving
equation \ref{x1}, but clearly, the origin of $x_{1}$ is arbitrary,
since the physics is translation-invariant, and does not depend on the
initial size of the dense region. The velocity of the interface, on the
other hand, is well defined:
\begin{equation}
\label{velocity}
  \dot{x}_{1}=\frac{a}{2}\sqrt{\frac{D}{t}}.
\end{equation}
Equation \ref{velocity} lends a physical meaning to the coefficient
$a$, and so we will refer to it as the `velocity coefficient'.

\begin{figure}
  \epsfysize=9cm
  \begin{center}
  \leavevmode\epsffile{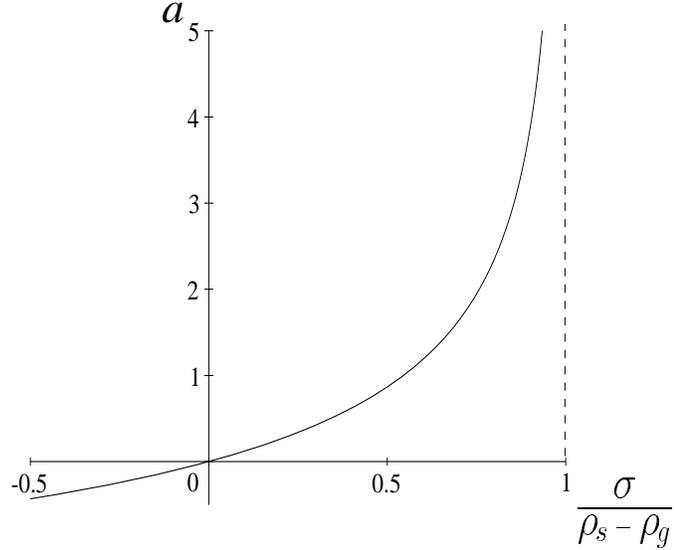}
  \caption{The velocity coefficient versus relative supersaturation.}
  \label{coeff}
  \end{center}
\end{figure}

  Equation \ref{answer} is plotted in figure \ref{coeff}, for $a$ as a
function of $\sigma/(\rho_{s}-\rho_{g})$. The validity of the model
extends to negative supersaturations, which result in evaporation of
the dense phase, and hence negative $a$. The curve plotted in figure
\ref{coeff} has no special features in the negative quadrant. As the
relative supersaturation tends to negative infinity, the behaviour of
the velocity coefficient is given by
\begin{displaymath}
  a \rightarrow -2 \sqrt{\mbox{\large ln}
  \left(\frac{-\sigma}{\rho_{s}-\rho_{g}}\right)}\,.
\end{displaymath}

\subsection{Linear and Non-Linear Regimes}
\label{regimes}

  For small relative supersaturation (close to the origin of figure
\ref{coeff}), equation \ref{answer} tends to a linear relation:
\begin{equation}
\label{linear}
  a \rightarrow \frac{2}{\sqrt{\pi}}
   \left(\frac{\sigma}{\rho_{s}-\rho_{g}}\right).
\end{equation}
From equation \ref{BC2}, we see that the flux {\em onto} the interface
is $j_{\rm cond.}=(\rho_{s}-\rho_{g})\,\dot{x}_{1}$. If we
substitute equations \ref{velocity} and \ref{linear} into this
expression, we see that the flux of material from the dilute phase
condensing onto the interface is
\begin{equation}
\label{flux}
  j_{\rm cond.} \approx \sigma \sqrt{D/\pi t}
\end{equation}
in the linear regime. This is independent of $(\rho_{s}-\rho_{g})$, and
depends only on properties of the dilute phase. So the condensation
flux becomes independent of properties of the interface in this regime.
The physics behind this statement is as follows. The density at the
base of the interface is maintained at the constant value $\rho_{g}$
and this gives rise to a gradient, and hence a flux, in the
supersaturated dilute region. This gradient diminishes with time as
material is depleted from the region, but is enhanced by the motion of
the wall. It is clear from figure \ref{diff} that moving the wall to
the right  must accentuate the gradient. In the linear regime, the
motion of the wall is slow enough for this enhancement of the gradient
to be insignificant, and the flux onto the interface varies with time
as if the boundary to the diffusive region were fixed.
In figure \ref{coeff}, we see that the model gives rise to a
divergence (albeit unphysical) of the velocity coefficient at a
relative supersaturation of unity. The reason for this is now clear. At
this supersaturation, the non-linearity described above, whereby the
interface motion enhances the gradient, becomes extreme. The interface
can never deplete the `low' density region, and hence the gradient (and
therefore the flux) at its base remains infinite.

  Based on the exact solution of this simple model for interface motion
(specifically equation \ref{flux}), we proceed to study a more elaborate
situation.

\section{Metastable Phase Evolution between Two Competing Interfaces}
\label{competing}

  It was noted in section \ref{behaviour} that the onset of
non-equilibrium behaviour in colloid-polymer mixtures seems to be
connected with the appearance of a third minimum in the free energy
density, corresponding to a metastable liquid phase. Specifically, the
formation of amorphous, non-equilibrium material from nucleation-like dynamics
begins to occur at a supersaturation of the gas phase close to
the hidden gas-liquid binodal. We now develop a model for the phase ordering
dynamics in systems with a third minimum of this kind.

  In a system with such a three-well potential (like that illustrated
in figure \ref{fe}b), three different species of interfaces may exist
between regions of the various locally stable densities, during
intermediate- to late-stage ordering. If the gas, metastable liquid and
solid phases are denoted {\it g, l} and {\it s} respectively, then
\mbox{\it g--l} , \mbox{\it l--s}  and \mbox{\it g--s}  interfaces may move
through the system. If we use the `equilibrium interface' approximation, as
applied in section \ref{exact}, defining the densities above and below each
interface to be the coexistence values for equilibrium between the two
neighbouring phases, then a \mbox{\it g--s}  interface has the fixed densities
$(\rho_{g},\rho_{s})$, given by the double-tangent construction \cite{DeHoff93}
in figure \ref{fe}b. Accordingly, at a \mbox{\it g--l}  interface
the densities are $(\rho_{A},\rho_{B})$, and at a \mbox{\it l--s}  interface,
$(\rho_{C},\rho_{D})$, which are the metastable coexisting values given in
figure \ref{fe}b.

  Consider the idealized system introduced in section \ref{exact},
where a region of the dense solid has appeared in an otherwise
homogeneous, supersaturated system. The ambient density is within the
gas well of the bulk free energy density. As before, the dense region
will grow by depleting the supersaturated dilute region, but we can now
imagine two alternative ways in which the growth can proceed. {\em
Either} a \mbox{\it g--s}  interface propagates into the dilute phase, and the
system evolves as in section \ref{exact}, {\em or} \mbox{\it g--l}  and
\mbox{\it l--s}  interfaces form, and propagate separately; their motion being
controlled by diffusion, both in the ambient dilute region and in the
intervening region of metastable liquid. Once formed, a \mbox{\it g--s}
interface propagates stably with respect to small perturbations \cite{Evans97a}
and cannot easily be split into \mbox{\it g--l}  and \mbox{\it l--s}  parts.
Hence, which of these two modes of growth the system exhibits depends on which
was initiated at the nucleation stage. The criteria for which mode is initiated
during nucleation in a given system are unknown at present, although there has
been some conjecture \cite{Evans97a,Evanslet}. Let us accept that the
split-interface (\mbox{\it g--l}  and \mbox{\it l--s}) mode of evolution has
begun in our one-dimensional system, and calculate the subsequent growth
dynamics.

\subsection{Model System}
\label{model}

\begin{figure}
  \epsfxsize=12cm
  \begin{center}
  \leavevmode\epsffile{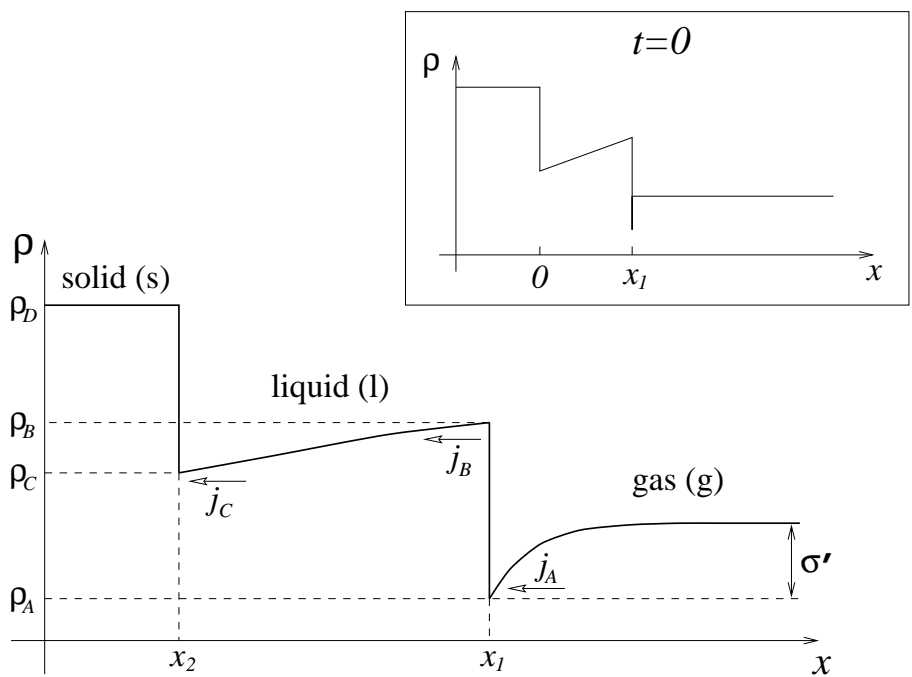}
  \caption{Graph of density $\rho$ against position $x$. The interface at
   $x_{1}$ separates the gas phase from the metastable liquid, and is locally
   equilibrated so that the discontinuity in the density is between the
   metastable binodal values $(\rho_{A},\rho_{B})$. The flux into (out of)
   the interface is $j_{A}$ ($j_{B}$). The flux $j_{C}$ of material of
   density $\rho_{C}$ impinges on the interface at $x_{2}$, which
   separates the metastable liquid from the solid region of uniform
   density $\rho_{D}$. The `ambient' ({\em i.e.\ }asymptotic) density of
   the gaseous phase is the supersaturated value $\rho_{A}+\sigma'$.
   Inset: The initial configuration of the system, with a uniform gradient
   between the domain walls, and a delta-function singularity in the gas
   phase at $x_{1}$.}
  \label{twowalls}
  \end{center}
\end{figure}

  The density profile of the system in question, at some time $t$, is
depicted in figure \ref{twowalls}. The \mbox{\it g--l}  interface is positioned
at $x_{1}(t)$, and the constant densities $\rho_{A}$ and $\rho_{B}$
immediately adjacent to it are marked, as are the fluxes $j_{A}(t)$ and
$j_{B}(t)$ in and out of the interface, which are defined in the
direction of the arrows, and are functions of time. The \mbox{\it l--s}
interface is at $x_{2}(t)$. The flux of material of density $\rho_{C}$ into
this interface is $j_{C}(t)$. There is no output flux, as the solid phase
has a uniform density $\rho_{D}$. Let us make the model as general and
physically realistic as possible by allowing different diffusion
constants in the different phases: $D_{1}$ in the gaseous phase, and
$D_{2}$ in the metastable liquid.
The supersaturated density at infinity is defined as $\rho_{A}+\sigma'$
in the figure. The `adjusted supersaturation' $\sigma'$ is distinct
from the supersaturation $\sigma$ used in section \ref{exact}. If the
ambient (asymptotic) density is $\rho_{\infty}$ then the
supersaturation is the deviation of this density from the equilibrium
value (the {\em stable} binodal), which is $\rho_{g}$ in this system,
{\em i.e.\ }$\sigma=\rho_{\infty}-\rho_{g}$. On the other hand, the
{\em adjusted} supersaturation is the deviation from the metastable
binodal, $\sigma'=\rho_{\infty}-\rho_{A}$.

  Clearly, conservation of matter at the interfaces leads to the equations
\begin{eqnarray}
  (\rho_{B}-\rho_{A})\dot{x}_{1} &=& j_{A}-j_{B}	   \label{conserv1} \\
  \mbox{and}\;\;\;(\rho_{D}-\rho_{C})\dot{x}_{2} &=& j_{C}. \label{conserv2}
\end{eqnarray}
Using these relations, we could proceed in the same manner as in
section \ref{solution}, but this time solving two coupled diffusion
equations. The diffusion equation for the gaseous region would have one
source, and that for the liquid region would have two. The strengths
and positions of the sources would be fixed by equations \ref{conserv1}
and \ref{conserv2}, and by fixing the constant densities $\rho_{A}$,
$\rho_{B}$ and $\rho_{C}$. Equation \ref{conserv1} would couple the two
equations. Proceeding in this manner, to try to find an exact solution
for the evolution of the system would lead to five coupled
integro-differential equations in five unknown functions ($x_{1}(t)$,
$x_{2}(t)$ and the strengths of three sources). It is not easy to spot
the solutions to this system of equations, as it was for equations
\ref{integ1} and \ref{integ2}. Instead we will make progress by
introducing some physically reasonable approximations.

\subsection{Approximations}
\label{apps}

  The first approximation is to decouple the \mbox{\it g--l} interface from
the gaseous driving region by using equation \ref{flux} for the input
current:
\begin{equation}
\label{approx1}
  j_{A} \approx \sigma' \sqrt{D_{1}/\pi t}.
\end{equation}
This becomes exact as $\sigma'\rightarrow 0$, and will presumably give
{\em qualitatively} meaningful results at higher supersaturations,
although with $\sigma'$ becoming some {\em effective} supersaturation,
deviating from the true value. We use $\sigma'$ rather than $\sigma$ in
equation \ref{approx1}, as the boundary condition
$\rho(x_{1})-\rho_{A}=0$ was used in its derivation.

  As a second approximation, let
\begin{equation}
\label{approx2}
  j_{B} \approx j_{C} \approx D_{2}\frac{\rho_{B}-\rho_{C}}{x_{1}-x_{2}},
\end{equation}
which says that the gradient in the liquid region is approximately
uniform. Intuitively, this seems to be a reasonable assumption. If we
imagine that, at some time into the evolution, the positions of the
interfaces could suddenly be frozen, then the liquid region would
subsequently relax exponentially towards a uniform gradient. (Compare
the related diffusion problem of the temperature profile in a
conducting rod with the ends held at different temperatures.) In this
sense, the liquid region is constantly in a state of relaxation towards
a uniform density gradient, for which equation \ref{approx2} applies.

  Let us define $\Delta$ to be the size of the metastable liquid
region; $\Delta\equiv x_{1}-x_{2}$. Then substituting equations
\ref{approx1} and \ref{approx2} in \ref{conserv1} and \ref{conserv2}
gives us the differential equation
\begin{equation}
\label{differential}
  \frac{d\Delta}{d(t/\tau)} = (t/\tau)^{-\frac{1}{2}} - \frac{\gamma}{2\Delta}
\end{equation}
with a constant
\begin{equation}
\label{time}
  \tau\equiv\frac{\pi}{D_{1}} \left(\frac{\rho_{B}-\rho_{A}}{\sigma'}\right)^2
\end{equation}
which has units of $\mbox{\em time}/\mbox{\em length}^2$, and a
dimensionless constant $\gamma$, which acts as an attractive coupling
between the walls, given by
\begin{equation}
\label{gamma}
  \gamma \equiv 2\pi \left(\frac{\rho_{B}-\rho_{A}}{\sigma'}\right)^2
   \frac{D_{2}}{D_{1}} (\rho_{B}-\rho_{C}) \left(
   \frac{1}{\rho_{B}-\rho_{A}} + \frac{1}{\rho_{D}-\rho_{C}} \right).
\end{equation}
The formula for $\gamma$ is quite easy to understand. It is a ratio of
quantities which drive the interfaces together (relating to diffusion
in the metastable phase) to those which drive them apart (relating to
the gaseous region). It is proportional to the ratio of the diffusion
constants, and to the difference in densities $(\rho_{B}-\rho_{C})$ which
drives flux through the liquid region. This difference is made
dimensionless by the factor following it, which is dominated by the
interface of smallest height. These quantities are divided by the
square of the relative supersaturation, which is responsible for
driving the \mbox{\it g--l}  wall away from the \mbox{\it l--s}  wall.

\subsection{Solution}

\begin{figure}[h]
  \epsfxsize=14cm
  \begin{center}
  \leavevmode\epsffile{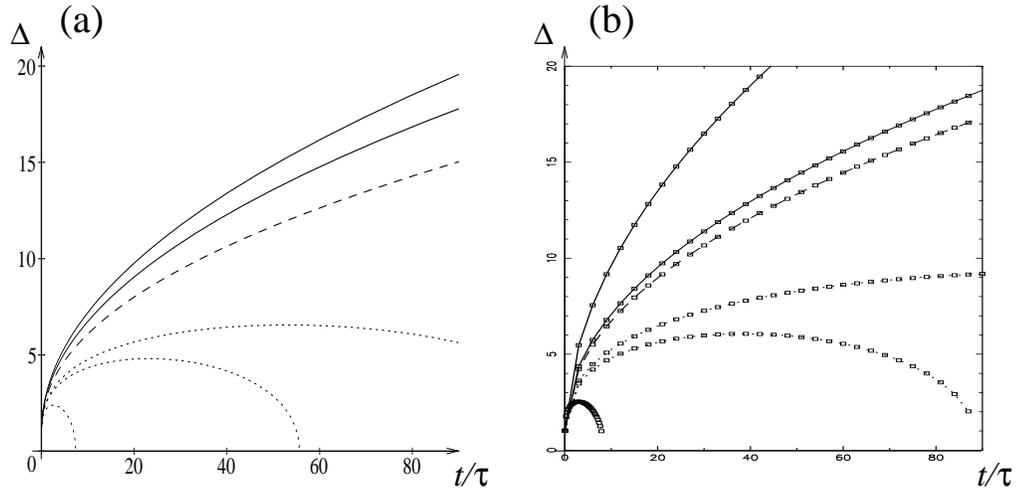}
  \caption{(a) Size of the metastable region $\Delta$ (as given in equations
   \protect\ref{bound}, \protect\ref{marginal} and \protect\ref{unbound})
   against time $t$ in units of the constant $\tau$, for various values of the
   coupling constant $\gamma$. Diffusively bound solutions are marked with
   a dotted line, for $\gamma=3$, $\gamma=2$ and $\gamma=1.8$. The
   marginal solution ($\gamma=1$) is dashed, and unbound solutions are
   plotted with solid lines, for $\gamma=0.8$ and $0.1$. In each case, the
   initial size of the metastable region, $\Delta_{0}$ is unity.
   (b) The same plots, produced by numerical solution, without the
   approximations, equations \protect\ref{approx1} and \protect\ref{approx2}.
   The same values of $\gamma$ are used as in (a).}
  \label{graphs}
  \end{center}
\end{figure}

  In equation \ref{differential}, $\gamma$ has a critical value of
unity, above which the solution may be expressed parametrically as
\begin{eqnarray}
  \sqrt{t/\tau} &=& \frac{\Delta_{0}}{\sqrt{\gamma-1}} \sin\theta\:
                    \exp\frac{\theta}{\sqrt{\gamma-1}}  \nonumber  \\
  \Delta &=& \frac{\Delta_{0}}{\sqrt{\gamma-1}}
             (\sin\theta +\sqrt{\gamma-1}\,\cos\theta)\:
               \exp\frac{\theta}{\sqrt{\gamma-1}}	    \label{bound}
\end{eqnarray}
for values of the parameter $\theta$ in the range
$0\leq\theta\leq\pi-\arctan\sqrt{\gamma-1}$. Here, $\Delta_{0}$ is the
initial size of the metastable region. A critical value of $\gamma$
implies (from equation \ref{gamma}) a critical value $\sigma_{c}$ of
the adjusted supersaturation. The condition $\gamma>1$ corresponds to
$\sigma'<\sigma_{c}$. We see that a graph of $\Delta$ versus $\sqrt{t}$
is an affine deformation of a logarithmic spiral, and the restricted
domain of $\theta$ gives a branch thereof in the first quadrant. Hence
the metastable region has a finite lifetime, since $\Delta$ decays to
zero at some positive value of $t$. We will refer to these solutions as
`diffusively bound', to distinguish from the tighter binding due to
curvature energy in the Cahn-Hilliard model, which prevents a single
\mbox{\it g--s}  interface splitting into \mbox{\it g--l}  and \mbox{\it l--s}
parts \cite{Evans97a}. Furthermore, there is no solution for $\Delta_{0}=0$.
Hence, for $\sigma'<\sigma_{c}$, the flux of condensation cannot, even
momentarily, separate the \mbox{\it g--l}  from the \mbox{\it l--s}  wall, if
they are initially together. The diffusively bound solutions
for $\Delta$ as a function of
$t/\tau$ are plotted with dotted lines in figure \ref{graphs}a for
various values of $\gamma$, with $\Delta_{0}=1$. Notice that the
gradients of the curves are infinite where they meet each axis. The
infinite gradient at $t=0$ arises from the infinite flux of condensing
material due to the delta singularity in the density at the base of the
\mbox{\it g--l}  interface. At $\Delta=0$, the metastable region decays
infinitely fast since the density gradient
$(\rho_{B}-\rho_{C})/\Delta$ diverges.
Of course, in a physical situation, the gradient of $\Delta(t)$ would
be flattened in both instances, since the densities
$(\rho_{A},\rho_{B})$ do not truly remain constant for a fast-moving
interface. Once the interfacial separation collapses to zero, the model
breaks down. In a real system, the interfaces would combine into a
single, stable \mbox{\it g--s}  interface, which would continue to advance.

  At the critical value $\gamma=1$, equation \ref{differential} has the
parametric solution
\begin{eqnarray}
  \sqrt{t/\tau} &=& \Delta_{0} (\phi+1)\: \exp\phi  \nonumber \\
  \Delta &=& \Delta_{0} \phi\: \exp\phi
\label{marginal}
\end{eqnarray}
and for $\gamma<1$ ({\em i.e.\ }above the critical supersaturation),
the solution is
\begin{eqnarray}
  \sqrt{t/\tau} &=& \frac{\Delta_{0}}{\sqrt{1-\gamma}}\sinh\phi\:
   \exp\frac{\phi}{\sqrt{1-\gamma}}
   \nonumber \\
  \Delta &=& \frac{\Delta_{0}}{\sqrt{1-\gamma}}
   (\sinh\phi +\sqrt{1-\gamma}\,\cosh\phi)\,
   \exp\frac{\phi}{\sqrt{1-\gamma}}
\label{unbound}
\end{eqnarray}
with $0\leq\phi<\infty$ in each case. Graphs of $\Delta$ versus
$t/\tau$, with $\Delta_{0}=1$, are plotted in figure \ref{graphs}a, with
solid lines for $\gamma<1$, and a dashed line for $\gamma=1$. These
solutions are not bound, {\em i.e.\ }$\Delta>0$ for all positive $t$
and $\Delta\rightarrow\infty$ as $t\rightarrow\infty$. So, above the
critical supersaturation, the flux of condensing material, although
always dwindling, is sufficient to separate the \mbox{\it g--l} from
the \mbox{\it l--s} 
interface, causing the metastable liquid phase to grow {\em ad
infinitum}. This is even true in the extreme case where $\Delta_{0}=0$,
when the solution becomes
\begin{equation}
\label{extreme}
  \Delta=\left(1+\sqrt{1-\gamma}\,\right) \sqrt{t/\tau}
   \;\;\;\;\;\;\;\;\mbox{for }\;\gamma\geq 1.
\end{equation}
Equation \ref{extreme} is also the limit of equation \ref{unbound} as
$t\rightarrow\infty$, so the size of the metastable region at late
times is independent of its initial value.

  To check the validity of equations \ref{approx1} and \ref{approx2}, we have
solved numerically for the system described in section \ref{model}
({\em without} the approximations made in section \ref{apps}). The results
plotted in figure \ref{graphs}b are for $D_{2}/D_{1}=0.1$. The variation of
$\gamma$ was controlled by varying $\sigma'$ only. These results
compare well with the approximate, closed-form solutions
in figure \ref{graphs}a. The same qualitative features (open/closed
trajectories) appear, and the critical value of $\gamma$ is close to unity.
The lifetimes of the bound solutions (which are very sensitive to the system
parameters) agree to within a few percent for systems not too close to
criticality. For the top-most trajectory in figure \ref{graphs}b,
$\sigma'/(\rho_{B}-\rho_{A})\approx 50\%$ so some deviation from linearity
(equation \ref{approx1}) is to be expected (see section \ref{regimes}).
We conclude that equations \ref{approx1} and \ref{approx2} are
quantitatively reasonable, and qualitatively very good, approximations.

  In summary, the asymptotic (late-time or long-distance) behaviour of
a metastable region, for a system in which a double interface has
formed early in the phase ordering, is as follows. For an ambient
supersaturation $\sigma'$ below the critical value
\begin{equation}
\label{critical}
  \sigma_{c} = (\rho_{B}-\rho_{A}) \sqrt{2\pi(\rho_{B}-\rho_{C})\left(
   \frac{1}{\rho_{B}-\rho_{A}}+\frac{1}{\rho_{D}-\rho_{C}}
   \right)\frac{D_{2}}{D_{1}}},
\end{equation}
there is no asymptotic behaviour, since the metastable region collapses
in a finite time, and subsequent evolution is via the ordinary
\mbox{\it g--s} mode
of interface propagation. Above the critical supersaturation, the
metastable phase grows with time, according to equation \ref{extreme}.
Notice, with reference to figure \ref{fe}b, that $(\rho_{B}-\rho_{C})$
is a measure of the metastability of the liquid phase. Hence at the
triple point, when the three wells have a single common tangent, the
critical supersaturation given by equation \ref{critical} goes to zero.
In other words, at the triple point, no supersaturation is required to
stabilize the liquid phase, so the physics is modelled successfully in
this respect.

  So far, we have concentrated on the size, $\Delta$, of the metastable
region, but the growth of the solid region is also of interest. For the
double-interface mode of evolution, the size of the solid region is
given, from equations \ref{conserv2} and \ref{approx2}, by
\begin{displaymath}
  x_{2}(t)=\left(\frac{\rho_{B}-\rho_{C}}{\rho_{D}-\rho_{C}}\right)
         D_{2}\int_{0}^{t}\frac{dt'}{\Delta(t')}.
\end{displaymath}
Substituting the asymptotic expression for $\Delta$ (equation
\ref{extreme}) into this formula, and using equation \ref{time} for
$\tau$, gives the result
\begin{displaymath}
  x_{2} = \frac{2\sqrt{\pi}}{(1+\sqrt{1-\gamma})} \frac{D_{2}}{\sqrt{D_{1}}}
           \left(\frac{\rho_{B}-\rho_{C}}{\rho_{D}-\rho_{C}}\right)
           \left(\frac{\rho_{B}-\rho_{A}}{\sigma'}\right) \sqrt{t}.
\end{displaymath}
Notice that this is {\em inversely} proportional to the (adjusted)
supersaturation. It is interesting to compare this with the size of a
solid region produced by normal interface motion ({\em i.e.\ }by a
single \mbox{\it g--s}  interface). To compare like with like, we should use the
linearized velocity coefficient (equation \ref{linear}, with
$(\rho_{B}-\rho_{A})$ replaced by $(\rho_{s}-\rho_{g})$ for the
interface height), together with equation \ref{x1}. Denoting the size
of the solid region produced by {\em single} interface motion by
$x_{\rm sing.}$, we find
\begin{displaymath}
  x_{\rm sing.}= \frac{2}{\sqrt{\pi}}\left(
   \frac{\sigma}{\rho_{s}-\rho_{g}}\right)\sqrt{D_{1}t}.
\end{displaymath}
Let $\zeta$ be the ratio of the size of the solid region produced by
the double-interface mode of growth, $x_{2}$, to that produced by
normal growth, $x_{\rm sing.}$. Then
\begin{displaymath}
  \zeta = \frac{\pi}{(1+\sqrt{1-\gamma})}
           \left(\frac{\rho_{B}-\rho_{A}}{\sigma'}\right)
           \left(\frac{\rho_{s}-\rho_{g}}{\sigma}\right) \frac{D_{2}}{D_{1}}\,
           \frac{(\rho_{B}-\rho_{C})}{(\rho_{D}-\rho_{C})}.
\end{displaymath}
There is a distinct similarity between this formula, and the expression
for $\gamma$ (equation \ref{gamma}). If we approximate
$(\rho_{D}-\rho_{A})+(\rho_{B}-\rho_{C})$ by $(\rho_{s}-\rho_{g})$,
which we see, from figure \ref{fe}b, is usually a good approximation,
then we find
\begin{displaymath}
  \zeta \approx \frac{\gamma}{2(1+\sqrt{1-\gamma})} \frac{\sigma'}{\sigma}.
\end{displaymath}
Since $0<\gamma<1$ for the split-interface mode of growth, and also
$\sigma'<\sigma$, it follows that
\begin{equation}
\label{inequality}
  \zeta < \frac{\gamma}{2}
\end{equation}
and that, well above the critical supersaturation, $\zeta\approx\gamma/4$.

  Let us recapitulate the properties of the parameter $\gamma$. It
appeared in equation \ref{differential} as an attractive coupling
between the \mbox{\it g--l}  and \mbox{\it l--s}  interfaces. It is the
ratio of properties of
the liquid phase, tending to attract the interfaces, to properties of
the gaseous phase, tending to separate them. It characterizes the
classes of solutions of equation \ref{differential}, being greater than
unity for diffusively bound solutions, and less than unity for unbound
solutions. Finally, we see in equation \ref{inequality} that it gives
an upper bound (and an order of magnitude) for the ratio of growth
rates of the solid phase in the split and normal modes of growth. Since
$\gamma<1$ when the split mode occurs, equation \ref{inequality} shows
that this mode of evolution {\em suppresses} the growth of the solid.

\begin{figure}
  \epsfxsize=6cm
  \begin{center}
  \leavevmode\epsffile{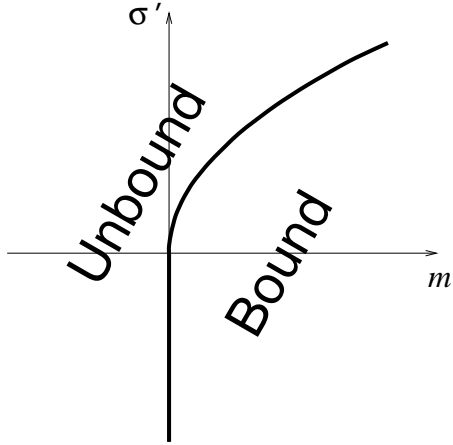}
  \caption{The regions of metastability $m\equiv\rho_{B}-\rho_{C}$ and adjusted
   supersaturation $\sigma'$ for which interfaces are diffusively bound or
   unbound.}
  \label{binding}
  \end{center}
\end{figure}

  Note that, in figure \ref{fe}b, the metastability of the middle well leads to
the inequality $\rho_{B}>\rho_{C}$. If this liquid well were stable
({\em i.e.\ }below the double tangent to the outer two wells), the inequality
would be violated. Hence it is reasonable to define the {\em metatability}, $m$,
of the middle well to be $m\equiv\rho_{B}-\rho_{C}$. With reference to
equation \ref{critical}, we see that the boundary between regions of the
$(m,\sigma')$-plane, for which the \mbox{\it g--l}
and \mbox{\it l--s} interfaces are diffusively bound
or unbound, is of the form shown in figure \ref{binding}. A na\"{\i}ve
expectation would be for a boundary coincident with the vertical axis, but we
see that this is not the case. (For negative $\sigma'$, for which the liquid
phase {\em must} dissolve into the gas, we find solutions of
equation \ref{differential} are again given by equations \ref{bound},
\ref{marginal} and \ref{unbound}, but with different ranges of the parameters
$\theta$, and $\phi$, leading to closed trajectories whenever $\gamma>0$.)

\section{Discussion}
\label{discussion}

  Some non-equilibrium effects in colloid-polymer mixtures have been
reviewed in section \ref{behaviour}. We have remarked on the presence
of a well-defined non-equilibrium boundary in the phase diagram,
close to the theoretically calculated position of the `hidden'
gas-liquid metastable binodal. This metastable binodal is central to the
theoretical model developed in section \ref{competing}, where the position
of its low-density branch was denoted by $\rho_{A}$. A system
homogenized by shear melting to an ambient density of exactly
$\rho_{A}$ would, in the notation of section \ref{competing} have an
vanishing `adjusted supersaturation', $\sigma'=0$. In the model, this
value has the special significance that, for $\sigma'>0$, a region of
metastable liquid has an initial period of growth, before it collapses.
At lower densities, any liquid region that is nucleated will immediately
shrink. In experimental samples with a colloidal density above the
non-equilibrium line in the phase diagram, the growth of crystals is
observed to be suppressed. This happens in the model for
$\sigma'>\sigma_{c}$.

  In practice the lines $\sigma'=0$ (the metastable binodal) and
$\sigma'=\sigma_{c}$ may be experimentally indistinguishable, for two
reasons. Firstly, no very accurate theoretical prediction exists for the
position of the metastable binodal in the phase diagram.  Attempting to
make quantitative comparisons of time scales with experiment, using the
best available theories for the values of $\rho_{A}$, $\rho_{B}$,
$\rho_{C}$ and $\rho_{D}$, results in uncertainties of several hundred
percent in the value of $\gamma$. Secondly, the line $\sigma'=\sigma_{c}$ may
be very close to the metastable binodal if $\sigma_{c}$ is small. (From
equation \ref{critical}, this would occur if e.g.\ the middle well in the
potential were only slightly metastable, or if the diffusion constant were much
lower in the liquid than in the gas.) If $\sigma_{c}$ is small then, writing
$\gamma=(\sigma_{c}/\sigma')^2$ for $\rho_{B}>\rho_{C}$, we see that
$\gamma$ will decrease rapidly with increasing colloidal density, and
so, from equation \ref{inequality}, suppression of crystal growth will
be pronounced, at densities not far from the metastable binodal.

  It seems, then, that the model presented here provides a plausible
theoretical basis for the previous conjecture that the onset of
nonequilibrium behaviour in certain colloid-polymer mixtures is
associated with the presence of a `hidden' gas-liquid binodal. It is,
however, a greatly simplified and idealized picture, and we should
consider the ways in which it deviates from reality, and the
implications for the resultant interfacial dynamics.

  The model is one-dimensional, and therefore ignores surface tension.
This is justifiable since, once a region has grown considerably larger
than the critical nucleus size, surface tension has a negligible effect
on interfacial motion during the growth stage \cite{Tokuyama93}. It is
at these intermediate-to-late times that our model describes the
system. Dimensionality is also relevant to the time dependence of the
long-range diffusion. This will have a quantitative effect on the
predicted values of the critical supersaturation and the degree of
suppression of crystal growth {\em etc.,} but we may conjecture that
the qualitative features of the model's behaviour will extend to three
dimensions. The fact that our one-dimensional model does not explicitly
address intrinsically higher-dimensional geometric effects, such as the
Mullins-Sekerka instability \cite{Mullins63}, may not be
important. The model gives us the general rule of thumb that, above
a certain critical supersaturation, the crystalline regions (whatever shapes
they may be), which would normally grow in a two-well system, are
replaced, in the presence of a third well, by metastable liquid.
Subsequent to this growth stage, the metastable liquid is slowly
transmuted into the `correct' equilibrium phase {\em i.e.\ }crystal.

  Probably a more drastic simplification is the semi-infinite extent of
the `ambient' gaseous region in the model. In reality, there is more
than one nucleus of the solid phase. There will be some typical
inter-nucleus spacing $L_{\rm nuc.}$ in the system. The concentration
profile in the gaseous region has a characteristic length scale
$\surd(D_{1}t)$ so, when this is of order $L_{\rm nuc.}$, the nuclei
begin to influence each other. After this time, the effective
asymptotic supersaturation begins to fall, as the regions of depletion
of the gas phase around the nuclei begin to overlap. This
characteristic time to deplete the supersaturation of the gaseous phase
may be denoted $t_{\rm dep.}$($\sim L_{\rm nuc.}^2/D_{1}$). At this
point, the metastable region around each nucleus is of size
$\Delta(t_{\rm dep.})$, which is given in equation \ref{extreme}. The
remaining lifetime of the metastable regions may be calculated from
equations \ref{bound}, as the time for a region initially of this size
to collapse, in a system with adjusted supersaturation of zero. The
total lifetime of the metastable phase is thus of order $t_{\rm
dep.}/\gamma$. (Compare this with the na\"{\i}ve calculation for a
semi-infinite gaseous region, which gives a suppression of size of the
solid region by a factor of order $\gamma$, and hence a time-scale
factored by $1/\gamma^2$, rather than $1/\gamma$.) We note that, if
$\sigma_{c}$ is small, then the lifetime of the metastable phase,
$(\sigma'/\sigma_{c})^{2}\,t_{\rm dep.}$ grows rapidly with the ambient
density. Once the metastable regions have collapsed, and
\mbox{\it g--s} interfaces
form, the densities will diffusively readjust from $\rho_{A}$ and
$\rho_{D}$ to $\rho_{g}$ and $\rho_{s}$, in a relatively short time.

  This whole discussion assumes that all interfaces in the system are
undergoing the `split' mode of propagation ({\em i.e.\ }with
\mbox{\it g--l} and \mbox{\it l--s}  parts
not bound together by curvature, as discussed in section \ref{competing} and
reference \cite{Evans97a}). Any ``normal" ({\it
g--s}) interfaces initiated in the system during nucleation will lead
to the formation of large crystals on normal diffusive time-scales. Such
crystals are not observed experimentally above the non-equilibrium
boundary. Therefore, our model is a good candidate for the physics of
the non-equilibrium boundary {\em if}, for some reason, only split
interfaces are generated during nucleation above this boundary. Such a
scenario is not unreasonable. We have seen that diffusively bound
interfaces are unbound by a sufficiently large supersaturation. It
seems likely \cite{Evans97a} (and has in fact been observed in a preliminary
numerical study \cite{Evanslet}) that a \mbox{\it g--s}  interface (which
is stabilized,
or ``bound" by a contribution to the chemical potential of the form
$-\nabla^2\rho$) may be ``split" or ``unbound" in a manner analogous to
diffusive unbinding and, furthermore, that the critical supersaturation
to cause this is close in magnitude
to that calculated here. This conjecture is based on the fact that, in a model
that includes an extra $\nabla^2\rho$ term in the chemical potential to describe
curvature effects (such as the Cahn-Hilliard model), the curvature term
exactly balances the diffusive term in an equilibrium interface and
hence quantities calculated from it will, in the main, be of the same
order of magnitude as those calculated from the diffusive term only.

\section{Related systems}
\label{related}

The model we have investigated in this paper was originally suggested
by experimental observations in mixtures of {\it spherical} colloids
and non-adsorbing polymers of a substantially smaller size. The same
model may, however, be applicable to other experimental systems.

First of all, it has been pointed out recently by one of us that the
`hidden binodal' is probably significant for understanding the
crystallization of globular proteins \cite{poon}. The kinetic
predictions of this paper may therefore also be relevant in that context.

Furthermore, our model is probably directly relevant to mixtures of {\it
rod-like} colloids and non-adsorbing polymers \cite{henk}. The two possible
phase diagram topologies for this system are again those given in figures
\ref{fe}a and \ref{fe}b, but with the labels gas, liquid and solid replaced
by $I_{1}$, $I_{2}$ and $N$ (standing for isotropic phases 1 and 2, and
nematic phase). Once more, three-phase coexistence disappears when the
size of the polymer (relative now to the rod length) decreases below a
critical value. In this case (compare figure \ref{fe}b),
addition of sufficient polymer to a suspension of rods leads to slow
phase separation into an isotropic and a denser, coexisting
nematic phase, the latter being distinguished by strong birefringence.
Further addition of polymers,
however, brings about a different kind of behaviour --- a weakly
birefringent, `expanded' phase, which contains most of the rod-like
particles, separates out quickly \cite{bruggen}. It has been
suggested \cite{bruggen} that the suppression of isotropic-nematic phase
separation is due to the presence of a `hidden' isotropic-isotropic
binodal.

\begin{figure}[h]
  \epsfysize=9cm
  \begin{center}
  \leavevmode\epsffile{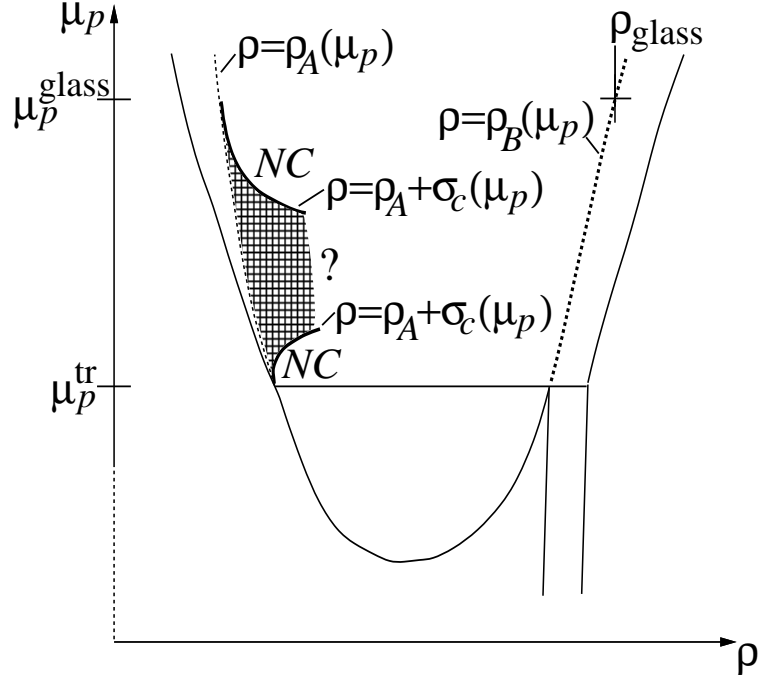}
  \caption{The phase diagram in figure \ref{pd}a redrawn schematically to
   emphasize the portion above the triple coexistence line at
   $\mu_{p}^{\rm tr}$. Here, within the equilibrium gas+solid coexistence
   region, the metastable gas-liquid binodal is the dotted lines
   $\rho=\rho_{A}(\mu_{p})$ and $\rho=\rho_{B}(\mu_{p})$. The bold
   curves indicate the likely positions of parts of the diffusive unbinding
   boundary $\rho=\rho_{A}+\sigma_{c}(\mu_{p})$. Near the triple coexistence
   line, the boundary meets the gas branch of the metastable binodal
   because the liquid well in the free energy density is only marginally
   metastable. At $\mu_{p}^{\rm glass}$, the liquid branch of the
   metastable binodal reaches the vitrification density, $\rho_{\rm
   glass}$. In this vicinity, the critical supersaturation curve
   is again close to the gas branch of the metastable binodal, now because
   of the vanishing diffusion constant in the liquid phase. $NC$ indicates
   regions where we expect disruption of crystallization, while
   normal gas-crystal coexistence is expected in the hatched area.}
  \label{meta}
  \end{center}
\end{figure}

Let us return to spherical colloid and polymer mixtures. In the preceding
sections, we have concentrated on polymer-colloid size ratios $\xi$
sufficiently small for no liquid phase to appear in the equilibrium phase
diagram (figure \ref{pd}b). Now consider the case
where the polymer is large enough to give rise to a thermodynamically stable
gas-liquid binodal, as appears in figure \ref{pd}a. Note that, in the
`gas+solid' region of figure \ref{pd}a, the form of $f(\rho)$ is as
sketched in figure \ref{fe}b, and hence the model set up in this paper is again
relevant. An expanded version of the phase diagram near
the triple line, now showing the metastable portions of the gas-liquid
binodal, is sketched in figure \ref{meta}. As the triple line is approached
from above ($\mu\rightarrow\mu_{p}^{\rm tr}$), where the liquid well is only
just metastable, the critical supersaturation  vanishes
($\sigma_{c}\rightarrow0$) because $(\rho_{B}-\rho_{C})\rightarrow0$; see figure
\ref{fe}b and equation \ref{critical}. At higher values of $\mu_{p}$, we expect
another regime where $\sigma_{c} \rightarrow 0$, this time due to the vanishing
of the diffusion coefficient in the liquid phase, $D_{2}$ (see equation
\ref{critical}). As $\mu_{p}$ increases, the density of the metastable liquid
phase, $\rho_{B}$ also increases; eventually $\rho_{B} \rightarrow \rho_{\rm
glass}$, the density at which the system vitrifies,  with $D_{2}
\rightarrow 0$.

We can thus speculate on the form of the boundary for diffusive unbinding,
$\rho=\rho_{A}(\mu_{p})+\sigma_{c}(\mu_{p})$, in the vicinity of the gas branch
of the metastable binodal, figure \ref{meta}. The prediction is that, for
moderate colloid densities, there should be a region of suppressed
crystallization immediately above the triple coexistence line (due to the very
marginal metastability of the liquid minimum in the free energy density),
followed, for higher $\mu_{p}$, by normal crystallization behaviour, and ending
up with crystallization suppression again at the highest values of $\mu_{p}$
(due to the vitrification of the metastable liquid phase). Experimentally, the
former non-equilibrium region has not been observed \cite{Ilett95}. A search for
this phenomenon is under way in our laboratory. However, the region of normal
crystallization has been reported, as has a non-equilibrium boundary at higher
$\mu_{p}$ \cite{Ilett95}, where it has been suggested \cite{nynke} that
vitrification of a dense phase does play a crucial role.

We speculate that the theoretical results contained in this paper should have
some relevance to a number of other complex fluid systems in which metastability
is known to play a key role in phase transformations, including the
crystallization of the `monotropic liquid crystal', poly-n-nonyl-4
4$^{\prime}$-biphenyl-2-cholroethane, via an intermediate, metastable
nematic phase \cite{heberer} and the crystallization of poly(phenylene
ether) in cyclohexanol, where deep quenches produces first a fluid-fluid
phase separation \cite{berghmans}. We acknowledge, however, that the
limitations of our model discussed in section \ref{discussion}, together with
the likelihood that latent heat may not be negligible in at least some
instances, necessitates further research before reaching firm conclusions on
its applicability.

\section{Conclusions}

  We have studied a simple model for diffusion-limited kinetics of
phase ordering in a system whose free energy density has a metastable
third well, at a density intermediate to the two equilibrium phases. In
such systems, we have found a mechanism whereby a region of the
metastable phase may grow {\em ad infinitum} at the expense of the
equilibrium dense phase, if the mean density of the system is above a
critical threshold. This behaviour appears to be consistent with the
non-equilibrium ordering dynamics and suppression of crystal growth
observed experimentally in colloid-polymer mixtures. In the
experimental system, as in the theoretical model, the onset of
anomalous behaviour occurs at a well defined density.

  The important lesson of this study (and the companion paper \cite{Evans97a})
is that any pair of concentrations, which can be linked by a double-tangent on
the graph of free energy density versus concentration, may give rise to an
interface in the evolving system. Although only the globally stable binodal
densities will coexist in the equilibrated system, local and transient
coexistence can occur between {\em any} binodal pairs of densities in a
system which has not yet discovered its global equilibrium state.
Hence, metastable phases cannot be overlooked when modelling phase
ordering. Indeed, their importance has long been accepted on an empirical level,
particularly in a metallurgical context \cite{Cahn68,DeHoff93}. Ostwald's
`rule of stages', for example \cite{Ostwald}, asserts that the transformation
from one stable phase to another proceeds via all metastable intermediates in
turn. We have, in the present paper, a rationale for the consideration of such
phases in a soft condensed matter setting.

\section{Acknowledgements}

  We thank M. E. Cates for his central involvement in discussion,
motivation and proof-reading. This work was supported by EPSRC grant number
GR/K56025. W. C. K. P. thanks the Nuffield Foundation for a science fellowship.

\appendix

\section{Evaluation of Integrals}
\label{integrals}

  Dividing equation \ref{integ1} by equation \ref{integ2}, and using
the appropriate functional forms for $x_{1}(t)$ and $s(t)$, and the
substitution $t'=u^2t$ to eliminate dimensionality from the integrands,
we find the relation between the velocity coefficient and the relative
supersaturation $\sigma/(\rho_{B}-\rho_{A})=I_{1}/I_{2}$ where
\begin{eqnarray*}
  I_{1} &\equiv& \int_{0}^{1} \frac{du}{\sqrt{1-u^2}}\,
   \exp-\frac{a^2(1+u)}{4(1-u)}  \\
   \mbox{and}\;\;\; I_{2} &\equiv&
   \int_{0}^{1} \frac{du}{(1-u)\sqrt{1-u^2}}\, \exp-\frac{a^2(1+u)}{4(1-u)}.
\end{eqnarray*} 
Making a further change of variables $v^2=(1+u)/(1-u)$ yields
\begin{eqnarray*}
  I_{2} &=& \int_{1}^{\infty} e^{-\frac{a^2v^2}{4}}dv
   = \frac{\sqrt{\pi}}{a} \,\mbox{erfc}\,\frac{a}{2}  \\
   \mbox{and}\;\;\; I_{1} &=& 2\int_{1}^{\infty}
   \frac{e^{-\frac{a^2v^2}{4}}}{1+v^2}dv
   = 2e^{\frac{a^2}{4}} \int_{1}^{\infty}
   \frac{e^{-\frac{a^2}{4}(1+v^2)}}{1+v^2}dv.
\end{eqnarray*}
We see that $I_{1}$ and $I_{2}$ are related by a derivative:
\begin{displaymath}
  \frac{d}{da}(I_{1} e^{-\frac{a^2}{4}}) = -aI_{2} e^{-\frac{a^2}{4}}.
\end{displaymath}
Hence $I_{1}$ is given by the indefinite integral
\begin{displaymath}
  I_{1} = -\sqrt{\pi}e^{\frac{a^2}{4}}\int
   e^{-\frac{a^2}{4}}\,\mbox{erfc}\,\frac{a}{2}\,da
\end{displaymath}
whose constant of integration is set by noting that $I_{1}$ vanishes as
$a\rightarrow\infty$. This integral is soluble by parts, using
\begin{displaymath}
  \int\!\,\mbox{erfc}\, x\, dx = x\,\mbox{erfc}\, x
   - \frac{e^{-x^2}}{\sqrt{\pi}} + \mbox{const.}
\end{displaymath}
The solution is
\begin{displaymath}
  I_{1} = \frac{\pi}{2} e^{\frac{a^2}{4}} (\,\mbox{erfc}\, \frac{a}{2})^2
\end{displaymath}
from which equation \ref{answer} follows.

\end{document}